\documentclass{iau}
\usepackage{graphicx}
\usepackage{color}

\title[The Void Galaxy Survey: Morphology and star formation properties of void galaxies] %% give here short title %%
{The Void Galaxy Survey: \\ Morphology and Star Formation Properties of Void Galaxies}

\author[Beygu et al.]   %% give here short author list %%
{Burcu Beygu$^1$, Kathryn Kreckel$^2$,Thijs van der Hulst$^1$, Reynier Peletier$^1$, Tom Jarrett$^3$, Rien van de Weygaert$^1$,
 Jacqueline H. van Gorkom$^4$, Miguel Arag\'{o}n-Calvo$^5$}

\affiliation{$^1$ Kapteyn Astron. Inst., Univ. Groningen, PO Box 800, 9700 AV Groningen, the Netherlands \\ email: {\tt beygu@astro.rug.nl} \\[\affilskip]
 $^2$MPIA, K\"{o}nigstuhl 17, 69117 Heidelberg, Germany\\[\affilskip]
 $^3$Dept. Astron., Univ. Cape Town, Private Bag X3, Rondebosch 7701, South Africa \\[\affilskip]
 $^4$Columbia University, MC 5246, 550 W120th St., New York, NY 10027, USA\\[\affilskip]
 $^5$University of California, Riverside, CA 92521, USA}
\pubyear{2014}
\volume{308}  %% insert here IAU Symposium No.
\pagerange{119--126}
\jname{The Zeldovich Symposium Genesis and Growth of the Cosmic Web}
\editors{Rien van de Weygaert, Sergei Shandarin,\\ Enn Saar \& Jaan Einasto, eds.}
\begin{document}

\maketitle

\begin{abstract}
We present the structural and star formation properties of 59 void galaxies as part of the Void Galaxy Survey (VGS). Our aim is to study in detail the physical properties 
of these void galaxies and study the effect of the void environment on galaxy properties. We use Spitzer 3.6$\rm{\mu m}$ 
and B-band imaging to study the morphology and color of the VGS
galaxies. For their star formation properties, we use $\rm{H\alpha}$ and GALEX near-UV imaging. We compare our results to a range of galaxies of 
different morphologies in higher density environments. We find that the VGS galaxies are in general disk dominated and star forming galaxies. Their star formation rates 
are, however, often less than 1 $\rm{M_{\odot}}$ $\rm{yr^{-1}}$. There are two early-type galaxies in our sample as well. In $\rm{r_{e}}$ versus $\rm{M_{B}}$ parameter space, VGS galaxies occupy the same space as dwarf irregulars 
and spirals.

\end{abstract}
\firstsection % if your document starts with a section,
%\section{Introduction}

\section{Voids and void galaxies}
Voids are a prominent aspect of the Cosmic Web (see \cite{weyplaten2011} for a 
review). Surrounded by elongated filaments, sheetlike walls and dense compact clusters, 
they have formed out of primordial underdensities via an intricate hierarchical 
process of evolution (\cite{shethwey2004,aragon2010,aragon2013,rieder2013}). Their 
diluted substructure and population of galaxies remain as fossils of the earlier 
phases of this void hierarchy. 

Within this context, the pristine environment of voids represents an ideal setting for 
the study of environmental influences on galaxy formation and evolution. Largely unaffected 
by the complexities and processes that modify galaxies in high-density environments, 
the characteristics of void galaxies are expected to provide information on the role 
of environment in galaxy evolution. 

A few aspects stand out immediately. The void galaxies are in general gas rich and display a 
substantial star formation activity. They also have rather low stellar masses, which relates to 
the finding by theoretical studies that the mass function of galaxies and halos in voids has 
shifteed considerably towards lower masses (see e.g. \cite{goldberg2004,aragon2007,cautun2014}). 
What still remains to be understood is the remarkable low abundance of dwarf galaxies in 
void interiors (\cite{peebles2001}). Also, we may wonder in how far the gas 
accretion and outflow history of void galaxies has resulted in systematically different objects 
(e.g. \cite{keres2005,hoeft2010}).  

In an attempt to obtain more insight into the properties of void galaxies, in this study we explore 
the morphology, structural characteristics and star formation properties of void galaxies in 
the Void Galaxy Survey (VGS). 

%%%%%% Figure %%%%%%%
\begin{figure}[h]
% \vspace*{-2.0 cm}
\begin{center}
 \includegraphics[width=0.85\textwidth]{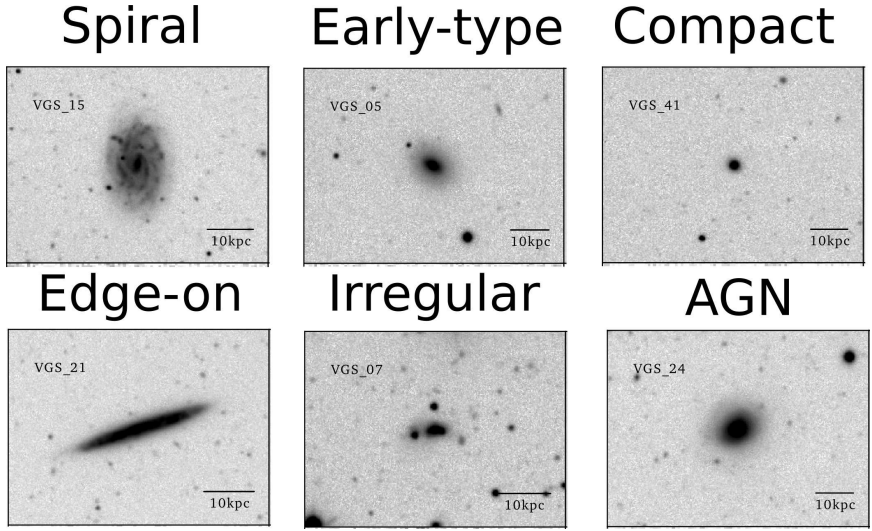} 
% \vspace*{-1.0 cm}
 \caption{VGS galaxy morphologies. The images are B-band images of 6 VGS galaxies. In each image, 
the black bar represents a physical scale of 10 kpc.}
   \label{fig1}
\end{center}
\end{figure}
%%%%%%%%%%%%%%%%%%%%%%%%%%%%%%%%%

\section{The Void Galaxy Survey}
 In order to fully study the effect of void environment on galaxy evolution and formation, one needs an
unbiased void galaxy sample found in a well defined void environment. The Void Galaxy 
Survey (VGS) (\cite{kreckel2011,weygaert2011,kreckel2012}, this volume) provides such a sample. This sample has been selected from the Sloan Digital Sky Survey Data Release 7 (SDSS DR7), 
using geometric and topological techniques for delineating voids in the galaxy distribution and identifying galaxies populating 
the central interior of these voids (\cite{schaap2000,platen2007,aragon2010a,kreckel2011}). The typical size of voids in our sample is 
on the order of 5 to 10 \textit{$h^{-1}$} Mpc in radius. 

The resulting sample of void galaxies is unbiased and largely independent of intrinsic galaxy properties 
(except for the the spectroscopic flux limit of 17.7 mag in the r-filter of the SDSS). 
The VGS galaxies have redshifts in the range $\rm{0.02 <  z < 0.03}$. 
They have an absolute magnitude in the range of $\rm{-20.4 < M_{r} < -16.1}$, colors in between 
$\rm{0.06 < g-r < 0.087}$ and a stellar mass $\rm{M_{*} < 3 \times 10^{10} M_{\odot}}$.
The Void Galaxy Survey aims to probe the color, morphology, star 
formation and gas content of the void galaxies. For this we observed 59 VGS galaxies in the 21 cm $\rm{H\textsc{i}}$ line, in 
$\rm{H \alpha}$ and in the optical B-band. In addition, we acquired GALEX near-UV data, as well as Spitzer 
3.6$\rm{\mu m}$ and WISE 22$\rm{\mu m}$ imaging. For some VGS galaxies we obtained CO(1-0) observations, which will 
form the starting point for a study of the relation between their star formation activity and their molecular and 
atomic gas content. 

%%%%%%%%%%%%%%Figure %%%%%%%%%%%%%%%%%%%%%%%
\begin{figure*}[h]
  \centering\includegraphics[width=0.8\textwidth]{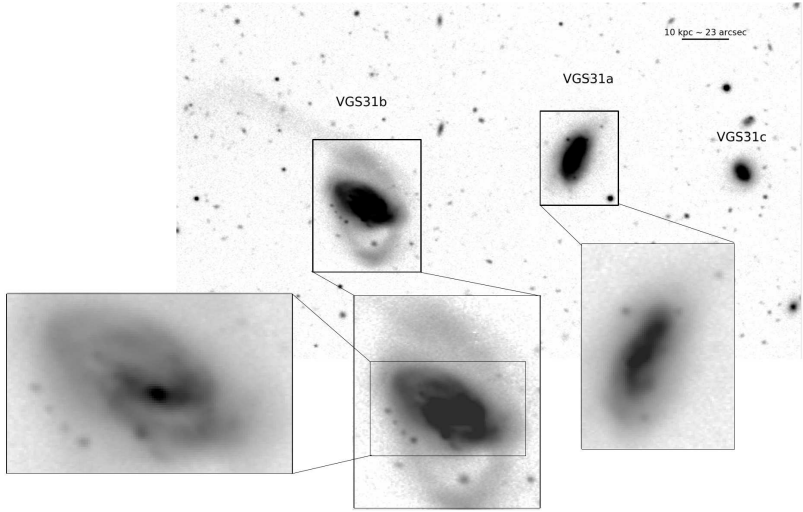}
  \caption{VGS\_31. R-band negative image of this configuration of three void galaxies aligned along a tenuous filament inside the void. 
From left to right: \textit{VGS\_31b}: The most remarkable member of the system, a \textit{Markarian} galaxy, 
has a tail and a ring. Close up images show the inner structures such as the bar.
\textit{VGS\_31a}: A disk galaxy with a bar structure. \textit{VGS\_31c}:
Smallest member of the system is optically undisturbed. The black bar on the top-right corner represents 10 kpc ($\sim$23$''$) (\cite{beygu2013}). } 
  \label{fig4}
\end{figure*}

%%%%%%%%%%%%%%Figure %%%%%%%%%%%%%%%%%%%%%%%%%%
So far we have completed the study of the $\rm{H\textsc{i}}$ properties of 55 VGS galaxies (see \cite{kreckel2012} for details ). 

\section{Morphology \& Structural Parameters}

The morphological classification of galaxies is rather complex and an accurate classification requires a 
more elaborate analysis than we are able to provide here. Therefore, instead of carrying out an absolute 
morphological classification, we seek to classify the morphology of the VGS galaxies in a general way, by eye. 
We find that the VGS galaxy sample mainly consists of disk galaxies with an occasional bar and spiral structure 
and sometimes small compact objects, irregulars and two early-types (one is an AGN). In figure~\ref{fig1} we 
display examples of the different morphological types of galaxies that we find in the VGS sample. Also, we 
find various peculiar galaxies, such as the dynamically distorted Markarian galaxy VGS\_31b in the filamentary 
VGS31 constellation (figure~\ref{fig4}, see \cite{beygu2013}).

\subsection{Structural parameters}
\begin{figure}[!htb] 
% \vspace*{-2.0 cm}
\centering
\includegraphics[width=0.99\textwidth]{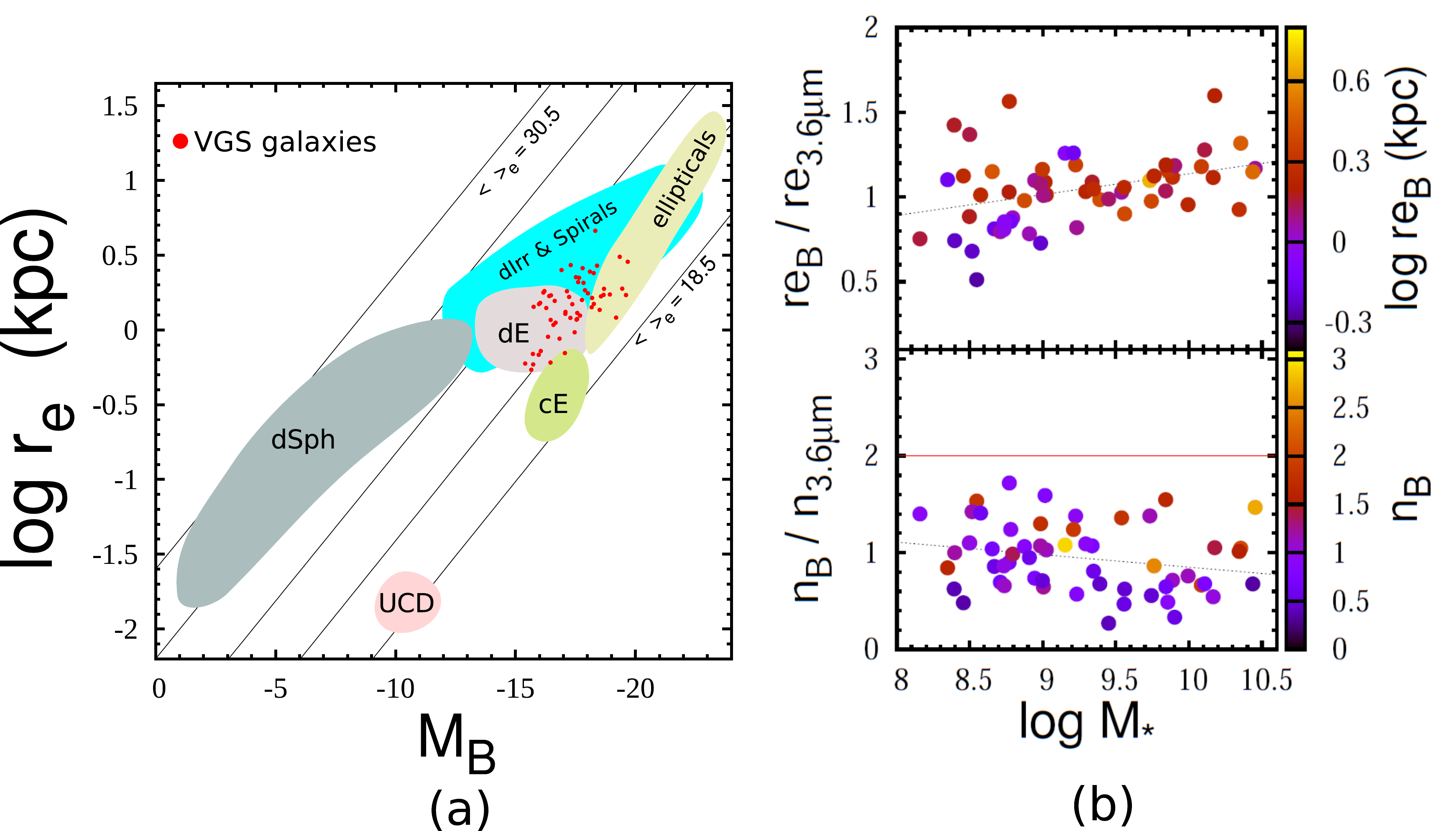}
% \vspace*{-1.0 cm}
 \caption{VGS galaxy structural parameters. (a): A sketch showing regions occupied by different types of galaxies and the VGS galaxies in the parameter space of the half-light radii $\rm{r_{e}}$ and B-band
absolute magnitude $\rm{M_{B}}$ adopted from \cite{mo2010}.
 VGS galaxies are shown as red dots. (b): The ratios of the $\rm{r_{e}}$ (top) and S\'{e}rsic indices (bottom) of the B-band and 3.6$\rm{\mu m}$ as a function 
 of stellar mass, $\rm{r_{e(B)}}$ and S\'{e}rsic index $\rm{n_{B}}$.}
   \label{fig2}
\end{figure}

The structural analysis of the VGS galaxies involves the fitting of S\'{e}rsic profiles to the light distributions 
and the determination of the characteristic size ($\rm{r_{e}}$) and surface brightness ($\rm{\mu_{e}}$) of the galaxies, 
along with their total luminosity. The concentration of (stellar) light is quantified by the S\'{e}rsic index \textit{n}. 
We have compared the determined structural parameter values for the VGS galaxies with the values of the same 
parameters for a wide range of different galaxies. Amongst these are giant early-type galaxies, dEs and late-type disk galaxies. 
Figure~\ref{fig2}a shows a sketch of where different types of 
galaxies are located in terms of their half-light radii and B-band absolute magnitudes. VGS galaxies (red dots), dIrr and spirals 
occupy the same space in $\rm{r_{e}}$ versus $\rm{M_{B}}$.

%%%%%%%%%%%%%%% Figure %%%%%%%%%%%%%%%%%%%%%%
%%%%%%%%%%%%%%%%%%% Figure %%%%%%%%%%%%%%%%%%%%%%

Figure~\ref{fig2}b shows the $\rm{r_{e,B}/r_{e,3.6}}$ (top panel) and the $\rm{n_{B}/n_{3.6}}$ ratios (lower panel) as a 
function of stellar mass, $\rm{r_{e,B}}$ and S\'{e}rsic index $\rm{n_{B}}$. The majority of the VGS galaxies have S\'{e}rsic 
indices \textit{n} $<$ 2 in both bands. This confirms that they are disk dominated. 

In figure~\ref{fig2}b, we see that $\rm{r_{e,B}/r_{e,3.6}}$ increases as a function of increasing stellar mass 
and as as well as a function of increasing $\rm{r_{e,B}}$. In other words, the smaller galaxies have a more 
concentrated light distribution at 3.6$\rm{\mu m}$ than in B. One explanation may be that in smaller galaxies the star formation activity is more 
concentrated towards the center than in the larger objects. Part of the effect is also caused by extinction. Larger 
galaxies generally contain more dust in their central regions. Dust affects the light in the B-band more 
than at 3.6$\rm{\mu m}$. This translates into a larger $\rm{(r_{e})_{B}}$ than $\rm{(r_{e})_{3.6}}$. 

%%%%%%%%%%%%%%% Figure %%%%%%%%%%%%%%%%%%%%%%%%%%%
\begin{figure*}[!htb]
\centering
  \centering\includegraphics[width=0.48\textwidth]{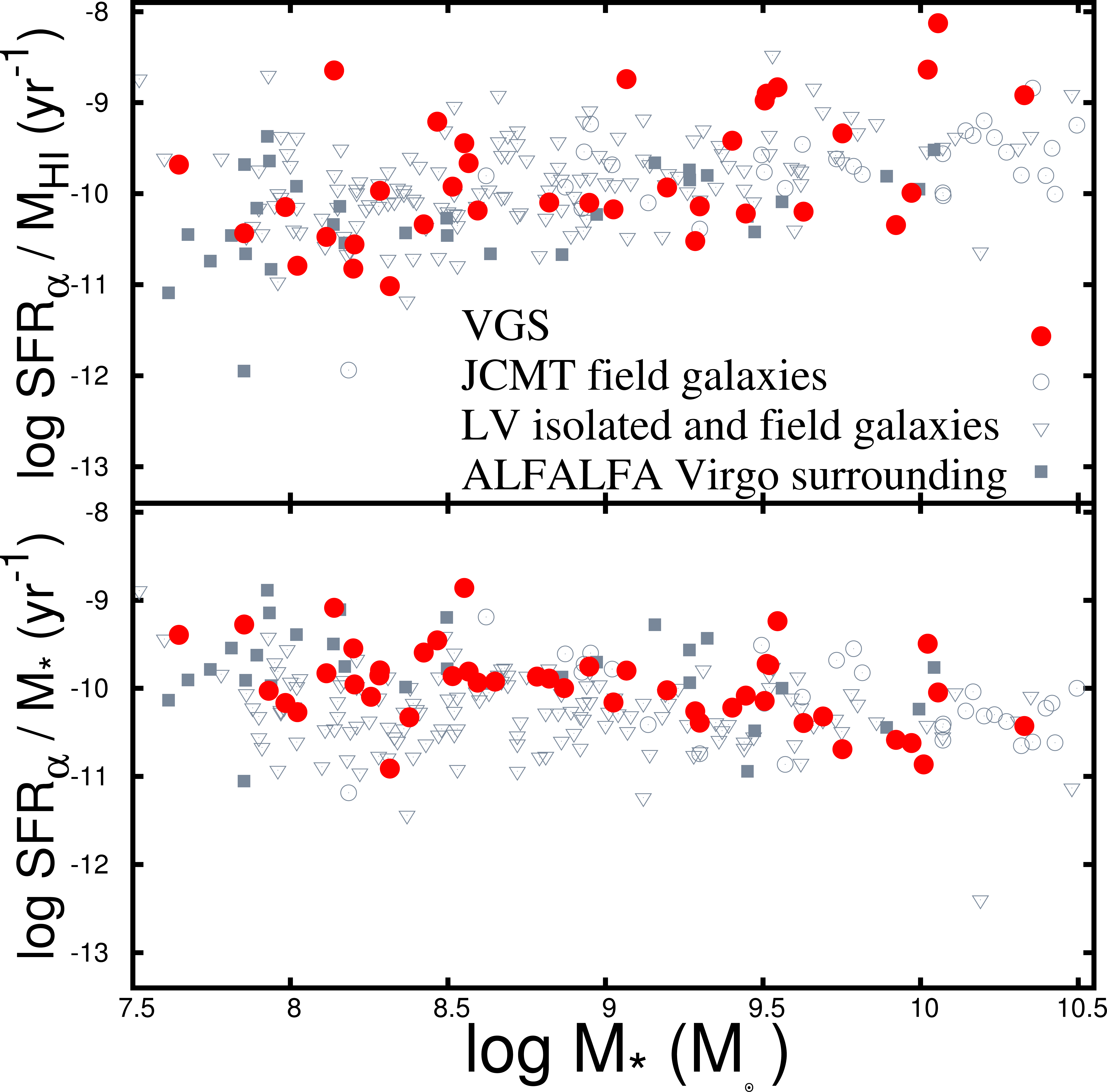} 
  \centering\includegraphics[width=0.49\textwidth]{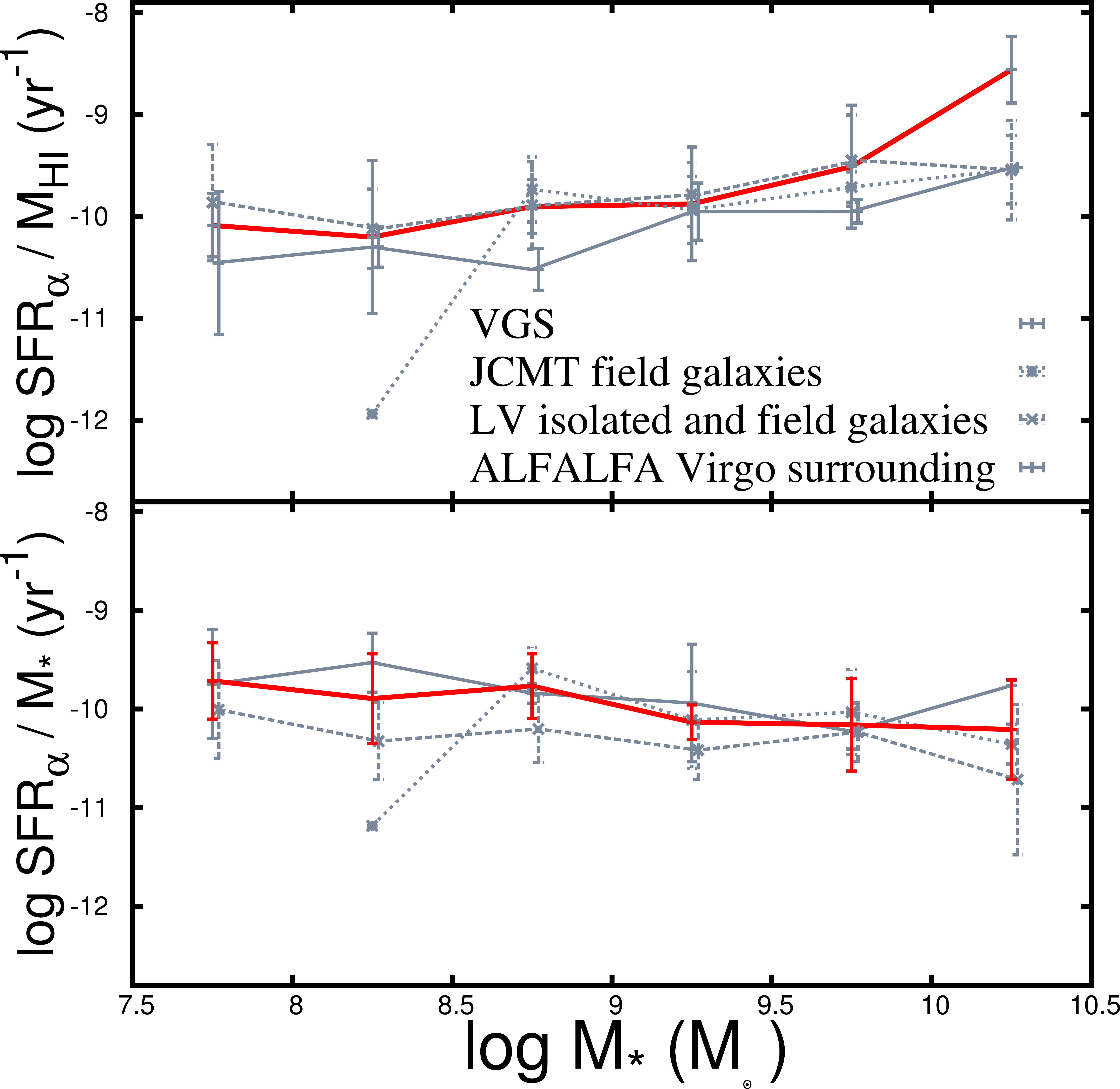}
\caption{Star formation properties of VGS galaxies. Top left: $\rm{SFR_{\alpha}}$/$\rm{M_{HI}}$ as a function of the stellar masss 
$\rm{M_{*}}$. VGS galaxies are indicated by red dots, the comparison sample galaxies are indicated by faint symbols. Bottom left: 
$\rm{SFR_{\alpha}}$/$\rm{M_{*}}$ as a function of stellar mass $\rm{M_{*}}$ for the sample sample of galaxies. 
Right panels: the average of the specific star formation parameters plotted in the corresponding lefthand panels. Note that there 
does not appear to be a significant difference between the VGS and the comparison sample galaxies. }
 \label{figure4}
\end{figure*}
%%%%%%%%%%%%%%%%%%%%% Figure %%%%%%%%%%%%%%%%%%%%%%%

\section{Star formation properties}

From $\rm{H \alpha}$ and near-UV imaging we may conclude that VGS galaxies are galaxies with a substantial star forming activity. 
Nonetheless, most of them appear to have star formation rates less than 1 $\rm{M_{\odot}}$ $\rm{yr^{-1}}$. The one exception with 
a considerably elevated star formation rate is the VGS\_31 (\cite{beygu2014}) system. 

We have compared the specific star formation ($\rm{SFR_{\alpha}/M_{*}}$) and star formation efficiencies - i.e. $\rm{SFR_{\alpha}/M_{HI}}$ - of the 
VGS galaxies to those of galaxies in average density regions (see \cite{beygu2014} for details). The latter belong to a sample 
of galaxies that consists of a combination of galaxies defined by three studies (\cite{gavazzi2012, sanchez2012} and \cite{karac2013}).  
These cover the same stellar mass range, as well as other properties, as the VGS galaxies. 

The left panel of figure 4 reveals a similar dependence of star formation efficiency and specific star formation rate 
on stellar mass for the VGS galaxies as for the galaxies from the control samples. As a function of stellar mass, the VGS and 
control sample galaxies show the same weak trends. In other words, we do not seem to detect a significant difference between the 
VGS galaxies and the control sample galaxies. 

\section{Conclusion}
The voids in our VGS sample do not appear to be populated by a type of galaxy specific for void environments. Voids mainly contain 
late-type galaxies of different morphologies. Only two VGS galaxies have an early-type morphology. While the void environment 
expresses itself in the low stellar mass and size of the galaxies (see e.g. \cite{aragon2007,cautun2014}), there is not evidence for 
star formation activity that deviates significantly from their peers with a similar mass in the higher denstiy filamentary and 
cluster-like environments of the cosmic web. It forms an indication for star formation to be a mainly self regulated process, not 
strongly influenced by the large scale environment. To understand this better, and to answer questions in how far the environment 
plays a role in initiating star formation, better theoretical understanding of the processes involved will be needed (see e.g. 
\cite{aragon2014}.

\end{document}